\documentclass[a4paper,11pt]{article}
\usepackage{jcappub} 
\usepackage{amsmath}
\usepackage{tikz}
\usepackage{cases}

\newcommand{\PTTpi}{P^{\rm{TT}}_\Pi}
\newcommand{\corr}[1]{\langle #1 \rangle}

\subheader{CERN--TH--2026--009, Imperial--TP--2026--PSG--1}

\title{\boldmath Gravitational waves from super-Hubble bubbles}

\author[a]{Pulkit S. Ghoderao,}
\author[a,b,c]{Arttu Rajantie}
\author[d]{and Dani\`{e}le A.~Steer}
\affiliation[a]{Abdus Salam Centre for Theoretical Physics, Imperial College London\\London SW7 2AZ, UK}
\affiliation[b]{Theoretical Physics Department, CERN\\1211 Geneva 23, Switzerland}
\affiliation[c]{Rudolf Peierls Centre for Theoretical Physics, University of Oxford\\Oxford OX1 3PU, UK}
\affiliation[d]{Laboratoire de Physique de l’ENS, Université Paris Cité, Ecole Normale Supérieure,\\ Université PSL,
Sorbonne Université, CNRS\\
75005 Paris, France}

\emailAdd{pulkit.ghoderao18@imperial.ac.uk; a.rajantie@imperial.ac.uk; daniele.steer@phys.ens.fr}

\abstract{
We consider a cosmological first-order phase transition in which bubbles of true vacuum nucleate during inflation but do not collide and percolate until the Universe has entered the radiation-dominated era. If the collisions take place soon after the end of inflation, the size of these bubbles can be significantly greater than the Hubble length.
This has two important consequences for the gravitational waves produced by the bubble collisions: 
First, as the peak frequency is determined by the comoving bubble radius, it can be well below the MHz frequencies typical for bubble collisions at the end of inflation. 
Second, when comparing collisions beginning at the same time, the amplitude of the gravitational waves is greatly enhanced relative to Hubble-sized and smaller bubbles.
Together, these two effects mean that gravitational waves produced by super-Hubble bubbles soon after the end of inflation can be within the observable frequency and amplitude ranges of LIGO and NANOGrav, as well as LISA and other future gravitational wave experiments. We demonstrate this with a simple illustrative model.
}

\begin{document}
\maketitle
\flushbottom
\section{Introduction}
\label{sec:Introduction}

One of the major science goals of the upcoming Laser Interferometer Space Antenna (LISA) experiment is to detect the stochastic gravitational wave background generated by cosmological first-order phase transitions~\cite{Caprini:2015zlo,LISA:2024hlh}. These occur when the Universe is in a metastable false vacuum and transitions to the true vacuum state, which has a lower vacuum energy density. The transition proceeds through nucleation of bubbles of true vacuum in the false vacuum background. The bubbles grow, collide, percolate and thus complete the transition. Both bubble collisions~\cite{Kosowsky:1991ua, Kosowsky:1992rz, Kosowsky:1992vn, Kamionkowski:1993fg} and subsequent non-equilibrium dynamics~\cite{Hogan:1986dsh, Hindmarsh:2013xza, Caprini:2009yp, Pen:2015qta} of the cosmological phase transition give rise to gravitational waves (GW). 

In most scenarios considered in the literature, the bubble size at collision is less than the Hubble length~\cite{Hogan:1983ixn}, or at most the same order of magnitude~\cite{Kobakhidze:2017mru}. In this paper, we show that there are scenarios in which the bubble size at collision can be much greater than the Hubble length, and that they can generate gravitational waves that are detectable at current and planned experiments.

We assume that the Universe is initially in the false vacuum with a negligible bubble nucleation rate. As it evolves forward in time, the bubble nucleation rate increases. Depending on the scenario, this can be because of the decreasing temperature or Hubble rate, or because of a change in the inflaton or other background fields.  

In the conventional scenario~\cite{Kosowsky:1992rz,Kamionkowski:1993fg, Apreda:2001us,Huber:2008hg,Caprini:2009fx, Jinno:2016vai,Hindmarsh:2016lnk,Konstandin:2017sat,LISACosmologyWorkingGroup:2022jok}, the nucleation rate increases monotonically, typically exponentially. As it crosses the threshold of one bubble nucleating per Hubble volume per Hubble time, multiple bubbles get formed within one Hubble volume and the transition to true vacuum is completed well-within a single Hubble time. In this case, the radius of bubbles at collision is typically a few orders of magnitude less than the Hubble length~\cite{Hogan:1983ixn}. 

In contrast, there are scenarios in which the nucleation rate remains low throughout, such that the threshold of one bubble nucleating per Hubble volume per Hubble time is never reached. This results in the physical separation between bubbles being greater than the Hubble length when they nucleate. In this case, the nucleated bubbles still grow, collide and complete the transition, but only after more than a Hubble time has elapsed since their nucleation. These are referred to as supercooled transitions~\cite{Kobakhidze:2017mru, Ellis:2018mja, Athron:2022mmm}, or in the case of inflation as slow transitions~\cite{Guth:1982pn, Barir:2022kzo}. In these cases,
the bubble radius at collision approaches the Hubble length due to the expansion of the background~\cite{Kobakhidze:2017mru}.

We consider a different scenario in which bubbles nucleate during inflation and collide shortly after inflation has ended. This can occur, for example, if there is a peak of the right magnitude in the nucleation rate. During inflation, the comoving Hubble length decreases, while the bubble radius continues to grow at the speed of light. As a result, the bubble radius can be significantly larger than the Hubble length when they collide after inflation has ended.

Super-Hubble bubbles like this, arising from inflation, have previously been considered in an astrophysical context to account for the voids observed in galaxy distribution~\cite{La:1990wy,Liddle:1991tr}. However, it was shown that super-Hubble bubbles violate Big Bang Nucleosynthesis (BBN) and Cosmic Microwave Background (CMB) constraints if they nucleate before the corresponding scales have left the horizon~\cite{Turner:1992tz}. This is not a problem in our scenario because the threshold at which a majority of the bubbles nucleate occurs after the BBN and CMB scales have left the horizon. Furthermore, the collision between super-Hubble bubbles occurs just after the end of inflation, much before the BBN or even the electroweak scale.   

In the conventional scenario, a flat spacetime treatment~\cite{Kosowsky:1991ua} is sufficient to estimate the gravitational wave signal expected from bubble collisions because their radius is much less than the Hubble length. For supercooled scenarios, the bubble radius is comparable to the Hubble length and thus a precise curved spacetime treatment corresponding to an expanding background becomes necessary~\cite{Zhong:2021hgo, Yamada:2025cfr}. In our case, where bubble radius is much greater than the Hubble length, we use a simpler analytical calculation to provide a conservative estimate of the gravitational wave signal. With this, we are able to clearly highlight the difference between our scenario and the conventional and supercooled ones.  

The stochastic gravitational wave background is often characterised by the spectral energy density 
\begin{equation}\label{eqn:GW spectrum definition}
    \Omega_{\rm{GW}}(f) = \frac{f}{\rho_{\rm crit}} \frac{d\rho_{\rm{GW}}(f)}{df},
\end{equation}
i.e. the energy density in GWs today $\rho_\text{GW}$, per logarithmic interval of observed GW frequency $f$, normalised by the critical energy density of the universe today $\rho_{\rm crit}$. A combined analysis of Big Bang Nucleosynthesis and Cosmic Microwave Background provides an upper bound on the gravitational wave spectrum over a large frequency range~\cite{Aggarwal:2025noe}
\begin{align}
    h^2\Omega_{\rm{GW}}(f) \lesssim 1.1 \times 10^{-6} \qquad \text{for} \qquad f \gtrsim 10^{-12} \text{Hz}.
\end{align} 
Here, $h$ is the Hubble rate today in units of $100$km\, s$^{-1}$Mpc$^{-1}$. Current experiments such as LIGO-Virgo-KAGRA (LVK) provide frequency-specific bounds e.g. $\Omega_{\rm{GW}}(f=25\rm{Hz})\leq 2.0\times 10^{-9}$ \cite{LIGOScientific:2025bgj}, whereas Pulsar Timing Arrays (PTA) have hinted at a detection of a signal within the frequency range $10^{-9}\mathrm{Hz}-10^{-7}\mathrm{Hz}$ at $h^2 \Omega_{\rm GW}= 10^{-10}-10^{-6}$~\cite{NANOGrav:2023hvm}.  Projections for upcoming GW experiments such as LISA~\cite{LISA:2024hlh} also have a frequency-dependent sensitivity, e.g. $\Omega_{\rm{GW}}(f=1\rm{mHz})\geq 3\times 10^{-11}$~\cite{Boileau:2022ter} for a signal from first-order phase transition to be detectable. 

The peak frequency $f^\text{peak}$ of gravitational waves produced in the conventional scenario as well as the supercooled scenario is of the order of the Hubble rate at the time when the bubbles collide, redshifted by the later expansion. When comparing collisions that begin at the same time, this implies that the peak frequency is related to the background temperature $T_c$ by the rule of thumb
\begin{equation}
\label{equ:fpeakthumb}
    f^\text{peak}\sim 10^{-7}\,\text{Hz}\,\frac{T_c}{\text{GeV}}.
\end{equation}
For example, transitions taking place at scales just after the end of inflation would produce gravitational waves in the MHz-range. Such a signal is undetectable with current observational methods.

However, for super-Hubble bubbles, we find that the GW peak
 gets shifted towards lower frequencies. Thus the rule of thumb (\ref{equ:fpeakthumb}) between frequency and background temperature at collision is broken in the case of super-Hubble bubbles. In addition, we find that the gravitational wave amplitude for super-Hubble bubbles is enhanced relative to smaller bubbles, when comparing collisions beginning at the same time. Thus, phase transitions occurring at inflationary energy scales can generate gravitational waves in the frequency and amplitude ranges probed by PTA, LISA, LIGO and future detectors.

This article is divided as follows: In Section~\ref{sec:Super-Hubble bubbles}, we begin by describing our scenario which gives rise to super-Hubble bubbles after inflation ends. Then we develop a simple formalism to estimate the gravitational wave spectrum from pairwise collisions of bubbles in a radiation background in Section~\ref{sec:Gravitational waves}. We apply this formalism to super-Hubble bubbles and estimate the peak frequency and amplitude of the GW signal arising from their collisions. Finally, in Section~\ref{sec:Illustrative model} we illustrate our scenario with the help of a specific model and show how a gravitational wave signal from super-Hubble bubble collisions can lie within the observed sensitivity ranges of current and planned GW experiments.

\section{Super-Hubble bubbles}
\label{sec:Super-Hubble bubbles}
We work in a spatially flat Friedmann-Lema\^itre-Robertson-Walker (FLRW) universe with metric
\begin{align}
\label{equ:FRWmetric}
    ds^2 = a^2(\eta) \left(-d\eta^2 + (\delta_{ij} + h_{ij}) dx^i dx^j \right), \qquad {i,j=1,2,3}
\end{align}
where $\eta$ is the conformal time, $a(\eta)$ the scale factor, and $h_{ij} \ll 1$ is the transverse and traceless tensor perturbation corresponding to gravitational waves.
We consider an inflationary period which ends at $\eta_\text{end}$ (when the scale factor and Hubble rate are respectively $a_\text{end}$ and $H_\text{end}$), and it is then followed by a radiation era. Assuming $H\approx$ constant during inflation,
\begin{equation}\label{eqn:scalefactor}
a(\eta)=
\begin{cases}
\displaystyle
\frac{a_\text{end}}{1+a_\text{end}H_\text{end}(\eta_\text{end}-\eta)},
&\text{when}~\eta\le \eta_\text{end},\cr
    a_\text{end} + a^2_\text{end} H_\text{end} (\eta - \eta_\text{end}), & \text{when}~\eta\ge\eta_\text{end},
\end{cases}
\end{equation}
and the comoving Hubble length ${\cal{H}}^{-1} \equiv (aH)^{-1}$ evolves as
\begin{equation}\label{eqn:comoving Hubble length}
 \mathcal{H}^{-1}(\eta)=
    \begin{cases}
        \displaystyle  \mathcal{H}^{-1}_\text{end}+\eta_\text{end}-\eta, & \text{when}~\eta\le \eta_\text{end},\cr
        \displaystyle \mathcal{H}^{-1}_\text{end}+\eta-\eta_\text{end}, & \text{when}~\eta\ge \eta_\text{end},
    \end{cases}
    \end{equation}
where $\mathcal{H}^{-1}_\text{end}$ denotes the comoving Hubble length at the end of inflation. 

\begin{figure}[t]
\centering
\begin{tikzpicture}[scale=0.66,line cap=round,line join=round]
\definecolor{cone}{RGB}{240,120,120}
\definecolor{bluec}{RGB}{0,70,255}
\coordinate (O)  at (0,1);      
\coordinate (L)  at (-5,6);     
\coordinate (T)  at (5,6);      
\coordinate (R0) at (10,1);      
\coordinate (R1) at (15,6);     
\draw[->,thick] (-4.5,0) -- (15.6,0)
node[below,align=center,yshift=-4pt] at (15.6,0)
{comoving\\distance};
\draw[->,thick] (0,-0.5) -- (0,7)
node[above] {$\eta$};
\draw[dashed] (-5.2,6) -- (15.2,6);
\draw[dashed] (-5.2,5) -- (15.2,5);
\draw[dashed] (-5.2,1) -- (15.2,1);
\node[right] at (15.2,6) {$\eta_c$};
\node[right] at (15.2,5) {$\eta_{\rm{end}}$};
\node[right] at (15.2,1) {$\eta_n$};
\draw[red,thick]
(L) -- (O) -- (T);
\fill[red!70,opacity=0.30]
(L) -- (O) -- (T) -- cycle;

\fill[cone,opacity=0.35]
(T) -- (R0) -- (R1) -- cycle;
\draw[red,thick]
(T) -- (R0) -- (R1);
\draw[dashed]
(10,-0.4) -- (10,6.3);
\coordinate (B) at (0.45,5);
\draw[blue,thick]
(B) -- (3,7.5);
\draw[blue,thick]
(B) -- (4.45,-0.0);
\node[blue] at (3.35,7.9)
{$\mathcal{H}^{-1}$};
\draw[<->,gray]
(0,6.3) -- (5,6.3);
\node[black] at (1,6.65)
{$R_c$};
\draw[<->,gray]
(0,-0.35) -- (10,-0.35);
\node[black] at (5,-0.7)
{$2R_c = \ell_n \mathcal{H}^{-1}_n$};
\end{tikzpicture}

\caption{Spacetime diagram illustrating the evolution of two bubbles \eqref{equ:Reta} (in red) and the comoving Hubble length $\mathcal{H}^{-1} \equiv(aH)^{-1}$~\eqref{eqn:comoving Hubble length} (in blue). The horizontal lines labelled by
$\eta_n, \eta_\text{end}$ and $\eta_c$ denote the conformal time at nucleation, end of inflation and collision respectively. The initial microscopic radius of bubbles at nucleation is ignored. 
}
\label{fig:Bubble wall evolution}
\end{figure}
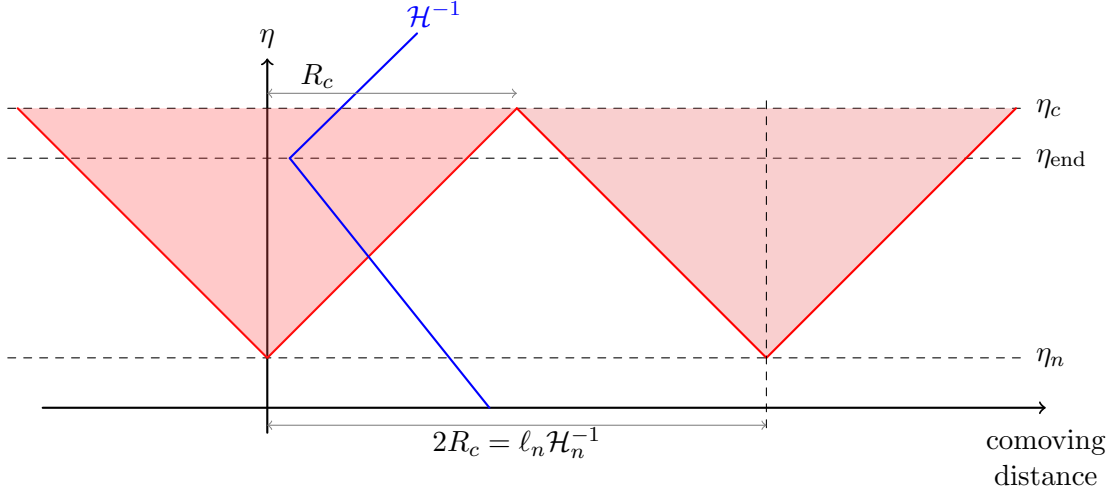

We assume that initially the Universe is in a metastable false vacuum state, which can decay to the true vacuum by bubble nucleation. The rate of this process is characterised by the nucleation rate per spacetime volume $\Gamma$.
After a bubble has nucleated, the bubble wall accelerates outwards because of the difference in vacuum energy between its interior and exterior. Neglecting any friction that would slow it down, its speed asymptotes to the speed of light. To a good approximation, we can therefore say that 
the 
comoving radius $R(\eta)$ of the bubble nucleated at conformal time $\eta_n$ evolves as
(see Figure~\ref{fig:Bubble wall evolution})
\begin{align}\label{equ:Reta}
    R(\eta) = \eta - \eta_n~.
\end{align}
If inflation were to continue forever, the conformal time would asymptote to a finite value, and therefore the comoving radius $R$ would also remain finite. However, because of the transition to radiation domination, the comoving radius $R$ grows without bound.

Bubble nucleation is a local, random process, and therefore bubbles nucleate uniformly across the whole Universe. As their comoving radius grows, they eventually collide with each other, producing gravitational waves~\cite{Kosowsky:1991ua}. The most commonly studied scenario~\cite{Kosowsky:1992rz} 
is when the nucleation rate grows exponentially with cosmic time, $\Gamma(t)\propto \exp(\beta t)$. Instead, 
we assume that the bubble nucleation rate peaks at a particular conformal time $\eta_n$, so that we can assume to a good approximation that all the bubbles nucleated at the same time.

To be more precise, the number density of bubbles per physical volume at cosmic time $t$ is given by
\begin{equation}
    n_\text{bub}(t)=\int_{-\infty}^t dt' \left(\frac{a(t')}{a(t)}\right)^3 \Gamma(t').
\end{equation}
In terms of the number of efolds, using $dN' = H(t') dt'$, this becomes
\begin{align}
    n_\text{bub}(N) = e^{3N} \int_{N}^{\infty} dN' e^{-3 N'} \frac{\Gamma(N')}{H(N')},
\end{align}
where the integrand 
\begin{equation}\label{eqn:gamma definition}
    \gamma(N)=e^{-3N}\frac{\Gamma(N)}{H(N)}
\end{equation}
gives the contribution from bubbles nucleated at $N$ efolds.
If $\gamma(N)$ is sharply peaked at a particular time $N_n$, then it is a good approximation to assume that all bubbles were nucleated at this same time. We label the corresponding conformal time as $\eta_n$.
We can then also define the parameter $\ell_n$ as the average distance between the bubble nucleation sites in units of the Hubble length,
and if we assume that $\Gamma(N)$ and $H(N)$ are slowly varying functions, we find
\begin{align}\label{eqn:ell_n in terms of Gamma}
    \ell_n =  n^{-1/3}_\text{bub}(N_n)~H_n \approx \left(\frac{\Gamma_n}{H^4_n}\right)^{-1/3},
\end{align}
where $\Gamma_n=\Gamma(\eta_n)$ and $H_n=H(\eta_n)$.

Thus, we assume that the bubbles nucleated at the same conformal time $\eta_n$ with an average separation $\ell_n$ in units of the Hubble length. They then collide at the time when their comoving radius has reached
\begin{equation}
    R_c=\frac{\ell_n}{2\mathcal{H}_n}.
    \label{eq:ellndef}
\end{equation}
According to eq.~(\ref{equ:Reta}), this happens at conformal time 
\begin{equation}
    \eta_c=\eta_n+R_c=\eta_n+\frac{\ell_n}{2\mathcal{H}_n}.
\end{equation}

Using the Friedmann equation, the bubble radius relative to the Hubble length, $\mathcal{H}R$, satisfies
\begin{align}\label{eqn:bubble radius evolution}
    \frac{d}{dN}(\mathcal{H}R) = 1 - \left(\frac{1+3w}{2}\right) (\mathcal{H}R),
\end{align}
where $N = \ln(a(\eta)/a_n)$ is the number of e-folds since nucleation and $w$ is the equation of state parameter of the dominant energy component ($w=-1, 1/3, 0$ during inflation, radiation and matter era respectively). 
Initially, at the time of nucleation, $\mathcal{H}R=0$, and therefore if the collision happens soon after nucleation, we will have $\mathcal{H}_cR_c\ll 1$, where $\mathcal{H}_c=\mathcal{H}(\eta_c)$. This means that the bubble radius is much shorter than the Hubble length.
At later times,
the solutions of eq.~(\ref{eqn:bubble radius evolution}) show that 
when $w>-1/3$, $\mathcal{H}R$ approaches a constant, $\mathcal{H}R\rightarrow 2/(1+3w)$. This means that after a sufficiently long period of radiation or matter domination, the bubble radius is always close to the Hubble length.
In contrast, during inflation when $w<-1/3$, the solution grows exponentially. The longer the bubble exists during inflation, the larger its radius becomes compared to the Hubble length. 
Hence, if the nucleation happens sufficiently early on during inflation and the collisions happen sufficiently soon after the end of inflation, 
then $\mathcal{H}_cR_c\gg 1$, i.e., the colliding bubbles are much larger than the Hubble length at the time of collision. We refer to such bubbles as
\emph{super-Hubble bubbles}.

Assuming that the collision takes place in the radiation era and taking the Hubble rate to be constant during inflation, it follows from eqs.~(\ref{eqn:scalefactor}) and (\ref{eqn:comoving Hubble length}) that the collision takes place when the scale factor is
\begin{equation}
\label{equ:ac}
    a_c=a_\text{end}\left[2+e^{N_n}\left(\frac{\ell_n}{2}-1\right)\right],
\end{equation}
where $N_n=\ln(a_\text{end}/a_n)$.
This means that the bubble separation parameter $\ell_n$ must satisfy
\begin{align}\label{eqn:ellnbound}
     \ell_n >   2 \left(1 - e^{-N_n}\right),
\end{align}
or otherwise the collision happens before inflation ends.
Furthermore, the bubble radius at collision in units of the Hubble length is
\begin{align}\label{eqn:infHcrc}
\mathcal{H}_c R_c = \left(
1-\frac{2-4e^{-N_n}}{\ell_n}
\right)^{-1}.
\end{align}
From this and eq.~(\ref{eqn:ellnbound}) it follows that $\mathcal{H}_cR_c$ can be as large as $e^{N_n}-1$. Therefore, as long as $N_n\gtrsim 1$, super-Hubble bubbles with $\mathcal{H}_cR_c\gg 1$ are possible and will arise if $\ell_n-2\ll 1$.

\section{Gravitational waves}
\label{sec:Gravitational waves}

\subsection{Gravitational wave production}

To calculate the produced gravitational wave spectrum, it is convenient to define the conformalised tensor perturbation $\tilde{h}_{ij} = a(\eta) h_{ij}$. In Fourier space, with $\vec{k}$ denoting the comoving momentum, it satisfies the equation
\begin{align}
\label{equ:htildeij}
    \frac{d^2\tilde{h}_{ij}}{d\eta^2} + \left(k^2 - \frac{1}{a}\frac{d^2 a}{d\eta^2}\right)\tilde{h}_{ij} = 16 \pi G a^3\Pi^{\rm{TT}}_{ij}(\vec{k},\eta),
\end{align}
where the source term $\Pi^{\rm{TT}}_{ij}$ is related to the stress-energy tensor $T_{ij}$,
\begin{equation}
\label{equ:PiTTdef}
\Pi^{\rm{TT}}_{ij}(\vec{k},\eta)=\frac{1}{a^2(\eta)}\Lambda_{ij,pq}(\vec{k})T_{pq}(\vec{k},\eta),
\end{equation}
through the transverse-traceless projection operator $\Lambda_{ij,pq}$ defined by 
\begin{equation}
\label{equ:Lambdadef}
    \Lambda_{ij,pq}(\vec{k}) \equiv \mathbb{P}_{ip}(\vec{k})\mathbb{P}_{jq}(\vec{k}) - \frac{1}{2} \mathbb{P}_{ij}(\vec{k})\mathbb{P}_{pq}(\vec{k}), \quad
    \mathbb{P}_{ij}(\vec{k}) \equiv \delta_{ij} - \frac{k_ik_j}{k^2}.
\end{equation}

We assume that the bubbles collide during radiation domination during which $d^2a/d\eta^2 = 0$ and the corresponding term in the equation of motion for the perturbation~\eqref{equ:htildeij} vanishes. Hence we do not have to pay special attention to the boundary between sub-Hubble ($k > aH$) and super-Hubble ($k < aH$) modes. In particular, even if the source term $\Pi_{ij}(\vec{k},\eta)$ involves a matter distribution evolving on super-Hubble scales, we can simply replace physical distances with comoving ones and physical time with conformal time, along with appropriate factors of $a$, to obtain the source term in radiation domination from its corresponding expression in Minkowski spacetime. Thus for bubble collisions, we shall take the Minkowski spacetime expression for the energy-momentum tensor traditionally used in the literature~\cite{Kosowsky:1991ua, Kosowsky:1992vn}, and readily obtain the corresponding expression in curved spacetime during radiation domination.

The gravitational wave power spectrum~\eqref{eqn:GW spectrum definition} can be re-written in terms of the comoving momentum using the definition
\begin{align}\label{eqn:physical and comoving frequency}
    f \equiv \frac{1}{2\pi} \frac{k}{a_0}
\end{align}
as
\begin{equation}
    \Omega_\text{GW}(k)=\frac{1}{\rho_\text{crit}}\frac{d\rho_\text{GW}}{d\log k}.
\end{equation}
The source for the gravitational wave spectrum that we are interested in is bubble collisions during the radiation era. It becomes active when the bubbles first collide, at conformal time $\eta_c$ and remains active until a majority of the spatial volume lies in the true vacuum, a time we denote by $\eta_f$.
Thus the GW spectrum is given by~\cite{Caprini:2018mtu}
\begin{align}\label{eqn:RDOmegaGW}
    \Omega_\text{GW}(k) = \frac{4}{\pi} \frac{1}{\rho_\text{crit}}\frac{G}{a^4_0} k^3 \int_{\eta_c}^{\eta_f} d\eta~a^3(\eta) \int_{\eta_c}^{\eta_f} d\eta'~a^3(\eta') \cos(k(\eta - \eta')) \PTTpi(k,\eta,\eta'),  
\end{align}
where $a_0$ denotes the scale factor today and $\PTTpi(k,\eta,\eta')$ is the power spectrum of the source ${\Pi}^{{\rm TT}}_{ij}(\vec{k},\eta)$ defined in terms of its two-point correlator as
\begin{align}
\label{eqn:power sepctrum for anisotropic stress}
    \corr{{\Pi}^{{\rm TT}}_{ij}(\vec{k},\eta) ~{{\Pi}^{{\rm TT}}_{ij}}^*(\vec{k}',\eta')} = \frac{(2\pi)^3}{4} \delta^3(\vec{k}-\vec{k}')  P^{{\rm TT}}_\Pi(k,\eta,\eta').
\end{align}

\subsection{Single collision}
\label{sec:single}

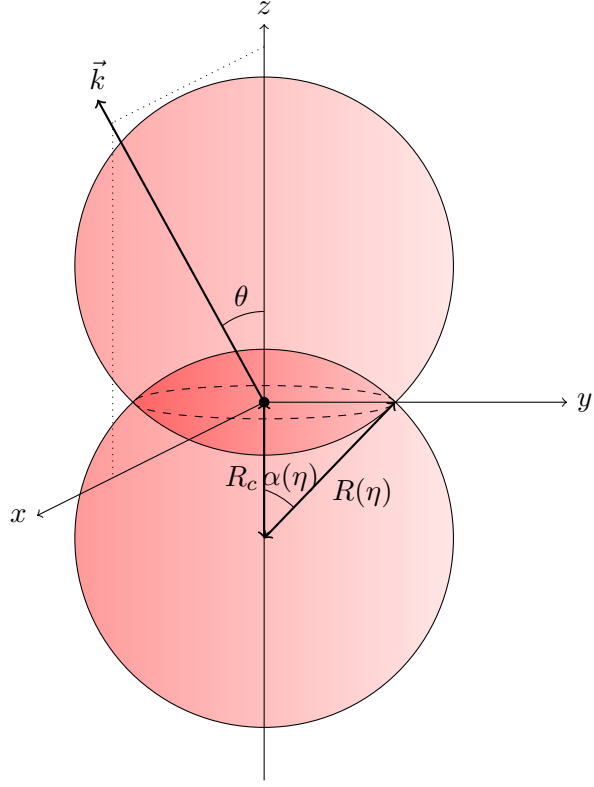
\begin{figure}[t]
\centering
\begin{tikzpicture}[scale=1]
    \coordinate (U) at (0,1.8);
    \coordinate (L) at (0,-1.8);
    \def\r{2.5}
\shade[left color=red!35,right color=red!10]
    (U) circle (\r);
\shade[left color=red!40,right color=red!10]
    (L) circle (\r);
\begin{scope}
    \clip (U) circle (\r);
    \shade[left color=red!70,right color=red!15]
        (L) circle (\r);
\end{scope}
\draw (U) circle (\r);
\draw (L) circle (\r);
    \draw[->] (0,0) -- (4,0) node[right] {$y$};
    \draw[->] (0,-5) -- (0,5) node[above] {$z$};
    \draw[->] (0,0) -- (-3,-1.5) node[left] {$x$};
    \draw[dashed]
        (-1.72,0)
        arc[start angle=180,end angle=360,
            x radius=1.72,y radius=0.22];
    \draw[dashed]
        (1.72,0)
        arc[start angle=0,end angle=180,
            x radius=1.72,y radius=0.22];
  \draw[dotted] (0,4.7) -- (-2,3.7);
    \draw[dotted] (-2,3.7) -- (-2,-1);

    \draw[->,thick]
        (0,0) -- (-2.2,4)
        node[above] {$\vec{k}$};
    \draw[<->,thick]
        (0,-1.8) -- (1.73,0);
    \draw[<->,thick]
        (0,0) -- (0,-1.8);
    \fill (0,0) circle (2pt);
    \draw
        (0,1.2)
        arc[start angle=90,end angle=128,radius=0.9];
    \node at (-0.3,1.4) {$\theta$};
    \coordinate (C) at (0.4,-1.4);
    \draw (C)
        arc[start angle=45,end angle=72,radius=1];
    \node at (0.35,-1) {$\alpha(\eta)$};
    \node at (1.3,-1.2) {$R(\eta)$};
    \node at (-0.3,-1) {$R_c$};
\end{tikzpicture}
   \caption{Schematic illustration of the collision between a pair of bubbles separated by a comoving distance $2R_c$ along their collision axis, here aligned along the $z$-direction. Also shown is the comoving momentum $\vec{k}$. The collision axis makes an angle $\theta$ with the comoving momentum.}
    \label{fig:Equal bubble collision}
\end{figure}

To calculate the power spectrum~(\ref{eqn:power sepctrum for anisotropic stress}), initially consider a collision between a single pair of equal-sized bubbles whose centres are separated by a comoving distance $2R_c$. A more complete picture involving multiple bubbles will be considered in the next section \ref{sec:multi}. The single pair setup is shown in Figure~\ref{fig:Equal bubble collision} where, without loss of generality, the collision is taken to be aligned with the $z$-axis. As the bubbles grow further, the overlap angle $\alpha(\eta)$ 
increases,
\begin{align}
    \alpha(\eta) = \cos^{-1}\left(\frac{R_c}{R(\eta)}\right).  
\end{align}
Owing to the cylindrical symmetry, we consider the comoving momentum $\vec{k}$ to lie in the $xz$ plane, and denote the angle between $\vec{k}$ and the collision axis by $\theta$, so that
$\vec{k} \equiv (k_x,k_y,k_z) = (k\sin\theta,0,k\cos\theta).$

We now follow ref.~\cite{Kosowsky:1992vn} and calculate the stress-energy tensor ${\overline{T}^i}_{j}$ for bubble pair collision oriented along the z-axis in the thin-wall and envelope approximations. However, ref.~\cite{Kosowsky:1992vn} did not include the effects of an expanding background since it only considered small bubbles which collide within a Hubble time. To account for the expanding background, we replace the distances and momenta from the Minkowski spacetime expression with comoving ones and include an additional factor of $a^2(\eta)$ that appears when lowering the spatial index $\overline{T}_{ij} = a^2 {\overline{T}^i}_{j}$. This yields
\begin{align}\label{eqn:T2envelope}
    \overline{T}_{ij}(k,\eta,\theta) = a^2(\eta)\left[\frac{\Delta\rho}{3}  R^3(\eta) \Theta_{ij}(k,\eta,\theta)
    -\delta_{ij}\mathcal{L}\right],
\end{align}   
where $\mathcal{L}$ is the Lagrangian density which  does not contribute to the source term in eq.~\eqref{equ:PiTTdef}, and
\begin{align}
     \Theta_{ij}(k,\eta,\theta) & =  
   e^{i k_zR_c} \int_0^{\pi-\alpha(\eta)} d\overline{\theta} \sin\overline{\theta} \int_0^{2\pi} \frac{d\overline{\phi}}{2\pi}~e^{i \vec{k} \cdot \vec{x}} \hat{x}_i \hat{x}_j
   \nonumber\\& \quad
    + e^{-i k_zR_c} \int_{\alpha(\eta)}^{\pi} d\overline{\theta} \sin\overline{\theta} \int_0^{2\pi} \frac{d\overline{\phi}}{2\pi}~e^{i \vec{k} \cdot \vec{x}} \hat{x}_i \hat{x}_j.
    \label{eq:thetas}
\end{align}
Here $\{\overline{\theta},\overline{\phi}\}$ are the spherical coordinates, while $\hat{x}$ denotes unit vectors in the $\{x,y,z\}$ directions. $\Delta\rho$ is the difference between the false and true vacuum energies, i.e., the exterior and interior of the bubbles at the collision time. To maintain consistency with the linearised gravity approximation we used to write the metric~\eqref{equ:FRWmetric}, we assume that throughout the collision
\begin{equation}
\label{eqn:kappa defn}
{\Delta \rho}\ll {\rho} ,
\end{equation}
where $\rho$ is the background energy density.

It is possible to perform the $\overline{\phi}$ integral analytically in terms of the Bessel functions of the first kind $J_{0,1,2}$ to yield~\cite{Kosowsky:1992vn}
\begin{align}\label{eqn:Envelope components}
    \Theta_{xx} &=  \int_{0}^{\pi - \alpha(\eta)} d\overline{\theta} \sin^3{\overline{\theta}} \cos\bigl(k_z R(\eta) \cos\overline{\theta} + k_z R_c\bigr) 
    \Bigl(J_0\bigl(k_x R(\eta) \sin \overline{\theta}\bigr) 
    - J_2\bigl(k_x R(\eta) \sin \overline{\theta}\bigr)  \Bigr),\nonumber\\
    \Theta_{yy} &= \int_{0}^{\pi - \alpha(\eta)} d\overline{\theta} \sin^3{\overline{\theta}} \cos\bigl(k_z R(\eta) \cos\overline{\theta} + k_z R_c\bigr) \Bigl(J_0\bigl(k_x R(\eta) \sin \overline{\theta}\bigr) + J_2\bigl(k_x R(\eta) \sin \overline{\theta}\bigr)  \Bigr),\nonumber\\
    \Theta_{zz} &= 2\int_{0}^{\pi - \alpha(\eta)} d\overline{\theta} \sin{\overline{\theta}} \cos^2\overline{\theta} \cos\bigl(k_z R(\eta) \cos\overline{\theta} + k_z R_c\bigr) J_0\bigl(k_x R(\eta) \sin \overline{\theta}\bigr), \nonumber\\
    \Theta_{xz} &= \Theta_{zx} = -2\int_{0}^{\pi - \alpha(\eta)} d\overline{\theta} \sin^2{\overline{\theta}} \cos\overline{\theta} \sin\bigl(k_z R(\eta) \cos\overline{\theta} + k_z R_c\bigr) J_1\bigl(k_x R(\eta) \sin \overline{\theta}\bigr) ,
\end{align}
with all other components vanishing. The source term due to this single collision $\overline\Pi^{TT}_{ij}(k,\eta,\theta)$, in this coordinate system, is then given by eq.~(\ref{equ:PiTTdef}) as
\begin{equation}
\label{equ:PibarTT}
    \overline\Pi^{TT}_{ij}(k,\eta,\theta)=\frac{1}{a(\eta)^2}\Lambda_{ij,pq}(\vec{k})\overline{T}_{pq}(k,\eta,\theta).
\end{equation}

\subsection{Multiple collisions}
\label{sec:multi}

In our scenario, space is filled with a large number of simultaneous bubble collisions with different, essentially random orientations. To further simplify the calculation, we consider only pairwise collisions, which is a good approximation at early times. Eventually it breaks when the likelihood of three or more bubbles overlapping becomes high. We cut off our calculation at that point, and because the later times make a positive additive contribution, this means that our result will be a conservative underestimate.

Let $m=1,2,\ldots, \mathcal{N}$ label the pairwise collisions whose midpoint is at $\vec{x}_m$. Then the total contribution from the source is the sum of contributions from all the individual pairs,
\begin{equation}
\label{equ:Pitot}
\Pi^{{\rm TT}}_{ij, \text{tot}}(\vec{k},\eta)=
\sum_m \Pi^{{\rm TT}}_{ij,m}(\vec{k},\eta) e^{i\vec{k}\cdot\vec{x}_m}, 
\end{equation}
where the exponential term arises from the corresponding term in eq.~\eqref{eq:thetas}. The two-point correlator of the source is then, 
\begin{eqnarray}\nonumber
    \left\langle{\Pi}^{{\rm TT}}_{ij, \text{tot}}(\vec{k},\eta) ~{{\Pi}^{{\rm TT}}_{ij, \text{tot}}}^*(\vec{k}',\eta')\right\rangle
    &=& \left\langle\sum_{m,m'} {\Pi}^{{\rm TT}}_{ij,m}(\vec{k},\eta) {\Pi}^{{\rm TT}}_{ij,m'}(\vec{k}',\eta') ~e^{i (\vec{k}\cdot \vec{x}_m - \vec{k}'\cdot \vec{x}_{m'})}
    \right\rangle
    \\
    &=& \sum_{m,m'} \left\langle{\Pi}^{{\rm TT}}_{ij,m}(\vec{k},\eta) {\Pi}^{{\rm TT}}_{ij,m'}(\vec{k}',\eta')\right\rangle 
    \left\langle e^{i (\vec{k}\cdot \vec{x}_m - \vec{k}'\cdot \vec{x}_{m'})}\right\rangle
\end{eqnarray}
where in the second line, the second factor is the average over pair positions $\vec{x}_m$, and
following the treatment in ref.~\cite{Caprini:2009fx}, we have 
made the approximation that the gravitational wave spectrum from a given pair (the first factor) does not depend on them. Furthermore using~\cite{Caprini:2009fx}, 
    \begin{align}
    \left\langle e^{i (\vec{k}\cdot \vec{x}_m - \vec{k}'\cdot \vec{x}_{m'})}\right\rangle =  \delta_{mm'}\frac{(2\pi)^3}{V}\delta^3(\vec{k}- \vec{k'}),
\end{align}
where $V$ is the comoving volume, we obtain an expression for the power spectrum~\eqref{eqn:power sepctrum for anisotropic stress} due to multiple bubble pairs as
\begin{align}
    \label{equ:PTT}
     P^{{\rm TT}}_{\Pi}(k,\eta,\eta')
    = \frac{4}{V}\sum_{m} \left\langle{\Pi}^{{\rm TT}}_{ij,m}(\vec{k},\eta) {\Pi}^{{\rm TT}}_{ij,m}(\vec{k},\eta')\right\rangle.
\end{align}

Assuming a statistically isotropic ensemble of bubble pairs, eq.~(\ref{equ:PTT}) is independent of the direction of the wave vector $\vec{k}$. Therefore the left-hand-side is written as a function of its magnitude $k$ only. On the right-hand-side we can choose the direction of $\vec{k}$ freely, and we choose to align it with the $z$-axis, i.e., $\vec{k}=k\hat{z}$. This isotropy also means that the ensemble average corresponds to simply integrating over the possible orientations of the bubble pairs. 

Let $\Pi^{\rm{TT}}_{ij}(k,\eta,\theta,\phi)$ denote the transverse-traceless source term due to a bubble pair oriented along the zenith and azimuthal angles $\theta$ and $\phi$, respectively. Then eq.~(\ref{equ:PTT}) becomes
\begin{align}
\label{equ:PTTint}
 P^{{\rm TT}}_{\Pi}(k,\eta,\eta') =  \frac{4}{V} \frac{\mathcal{N}}{4\pi}\int_0^{2\pi}d\phi\int_0^{\pi}d\theta\,\sin\theta\,
 {\Pi}^{{\rm TT}}_{ij}(k,\eta,\theta,\phi) {\Pi}^{{\rm TT}}_{ij}(k,\eta',\theta,\phi), 
\end{align}
where $\mathcal{N}$ is the total number of pairs. To obtain ${\Pi}^{{\rm TT}}_{ij}(k,\eta,\theta,\phi)$, we have to rotate $\overline{\Pi}^{{\rm TT}}_{ij}(k,\theta,\eta)$ calculated in eq.~(\ref{equ:PibarTT}), where the collision axis was oriented along the z-axis, to the present coordinate system where the collision axis is oriented along $(\theta,\phi)$. This can be achieved through the rotation matrix
\begin{align}
\label{equ:Qmatrix}
Q(\theta,\phi) = \begin{pmatrix}
-\cos\phi & \sin\phi & 0\\
-\sin\phi & -\cos\phi & 0\\
0 & 0 & 1
\end{pmatrix}
\begin{pmatrix}
\cos\theta & 0 & -\sin\theta\\
0 & 1 & 0\\
\sin\theta & 0 & \cos\theta 
\end{pmatrix},
\end{align}
as
\begin{equation}
    \Pi^{\rm{TT}}(k,\eta,\theta,\phi)=Q(\theta,\phi)\overline{\Pi}^{\rm{TT}}(k,\eta,\theta)Q^t(\theta,\phi),
\end{equation}
where the superscript $t$ indicates matrix transpose.

Using eqs.~(\ref{eqn:T2envelope}) and (\ref{equ:PibarTT}), the integrand in eq.~(\ref{equ:PTTint}) can now be written as
\begin{align}
      {\Pi}^{\rm{TT}}_{ij}(k,\eta,\theta,\phi) {\Pi}^{\rm{TT}}_{ij}(k,\eta',\theta,\phi) = \frac{\Delta\rho^2}{18} R^3(\eta) R^3(\eta') F(k,\eta,\theta) F(k,\eta',\theta),
\end{align}
where
\begin{align}\label{Fdefn}
   F(k,\eta,\theta) \equiv \Theta_{zz}(k,\eta,\theta) \sin^2\theta + \Theta_{xx}(k,\eta,\theta) \cos^2\theta - \Theta_{yy}(k,\eta,\theta) - 2 \Theta_{xz}(k,\eta,\theta) \sin\theta \cos\theta,
\end{align}
and the functions $\Theta_{ij}$ are given by eq.~\eqref{eqn:Envelope components}. Thus the power spectrum becomes
\begin{align}\label{eqn:power spectrum result}
     P^{\rm{TT}}_{\Pi}(k,\eta,\eta') =  \frac{\Delta\rho^2}{9} R^3(\eta) R^3(\eta') \frac{\mathcal{N}}{V}  \int_0^{\pi}d\theta\,\sin\theta\,F(k,\eta,\theta) F(k,\eta',\theta) .
\end{align}
This result for the power spectrum arising from multiple bubble pairs oriented randomly is equivalent to averaging over all possible orientations of the power spectrum arising from a single bubble pair and multiplying by the number density of pairs.

\subsection{Gravitational wave spectrum}

Substituting our expression for power spectrum arising from multiple pairs~\eqref{eqn:power spectrum result} in eq.~(\ref{eqn:RDOmegaGW}),
and writing the cosine in terms of an exponential we obtain
\begin{align}\label{eqn:OmegaGWraddom}
    \Omega_\text{GW}(k) = \frac{4}{9\pi} \frac{\Delta\rho^2}{\rho_\text{crit}}\frac{G}{a^4_0} \frac{\mathcal{N}}{V} k^3 \int_0^\pi d\theta\sin\theta~ \Bigg| \int_{\eta_c}^{\eta_f} d\eta~ e^{i k \eta} a^3(\eta) R^3(\eta)F(k,\eta ,\theta)\Bigg|^2.
\end{align}
Using the relation $a^2_\text{end}H_\text{end} = a^2_cH_c$ in radiation domination, the scale factor~\eqref{eqn:scalefactor} can be re-expressed as
\begin{align}
    a(\eta) = a_\text{end} + a^2_c H_c \bigl(R(\eta) - R_\text{end}\bigr) = a_c \left(1 + \mathcal{H}_cR_c \left(\frac{R(\eta)}{R_c} - 1\right)\right).
\end{align}
Thus in terms of the dimensionless combinations $\mathcal{H}_cR_c$, $kR_c$ and $u = R/R_c$, our expression for $\Omega_\text{GW}(k)$ becomes 
\begin{align}\label{eqn:OmegaGW with I}
 \Omega_\text{GW}(k) \sim  10^{-2} G\frac{\Delta\rho^2}{\rho_\text{crit}}\frac{a^4_c}{a^4_0}~a^2_cR^2_c~ I(kR_c,\mathcal{H}_cR_c),
\end{align}
where we have approximated the comoving number density of bubble pair collisions by $\mathcal{N}/V = \nu/R^3_c$, with $\nu \sim 10^{-1}$ and
\begin{align}\label{eqn:function I}
I(&kR_c,\mathcal{H}_cR_c) = (kR_c)^3\int_0^\pi d\theta\sin\theta~    \Bigg|\int_{1}^{u_f} du~e^{i kR_c u} u^3 \left(1 + \mathcal{H}_cR_c \left(u - 1\right)\right)^3  F(k,\eta,\theta)\Bigg|^2.
\end{align}
Note that $F(k,\eta,\theta)$ is readily expressed in terms of $u$ using eq.~\eqref{eqn:Envelope components}.
Here we have cut off the upper limit of the integration at $u=u_f$, because when $R\gg R_c$ the assumption that only pairwise bubble collisions contribute is no longer justified. In principle, one could include multi-bubble collisions, but that would make the calculation considerably more difficult. In any case, the contribution from larger $u$  is only going to increase the gravitational wave amplitude, and therefore cutting off the integral gives us a reliable conservative estimate. In what follows, we will set $u_f=2$, but we shall quantify the dependence of our result on that choice in Figure~\ref{fig:XivHcrc}.

Using the Friedmann equation $\rho_c = 3H^2_c/8\pi G$, the GW spectrum~\eqref{eqn:OmegaGW with I} can be rewritten as
\begin{align}
 \Omega_\text{GW}(k) \sim 10^{-3} \frac{a^4_c}{a^4_0} \frac{\rho_c}{\rho_\text{crit}} \frac{\Delta\rho^2}{\rho^2_c} (\mathcal{H}_cR_c)^2~I(kR_c,\mathcal{H}_cR_c).
\end{align}
The pre-factor has a standard value~\cite{Dodelson:2020bqr}
\begin{align}
   \frac{a^4_c}{a^4_0} \frac{\rho_c}{\rho_\text{crit}} \approx \frac{4.15 \times 10^{-5}}{h^2}.
\end{align}
Thus we find the gravitational wave spectrum generated by collision of bubbles including curved spacetime effects in radiation domination to be 
\begin{align} \label{eqn:resultOmegaGW}
 h^2\Omega_\text{GW}(k) \sim  10^{-7} \frac{\Delta\rho^2}{\rho^2_c} (\mathcal{H}_cR_c)^2~I(kR_c,\mathcal{H}_cR_c).
\end{align}

\subsection{Peak frequency and amplitude}
\label{sec:Peak frequency and amplitude}
All the momentum dependence of the gravitational wave spectrum~\eqref{eqn:resultOmegaGW} lies in the function $I(kR_c,\mathcal{H}_cR_c)$~\eqref{eqn:function I}. The parameter $\mathcal{H}_cR_c$ represents the size of bubbles at collision as compared to the Hubble length. Based on it, we can separate our discussion into sub-Hubble $\mathcal{H}_cR_c \ll 1$ and super-Hubble $\mathcal{H}_cR_c \gg 1$ regions.

\begin{figure}
    \centering
    \includegraphics[scale=0.3]{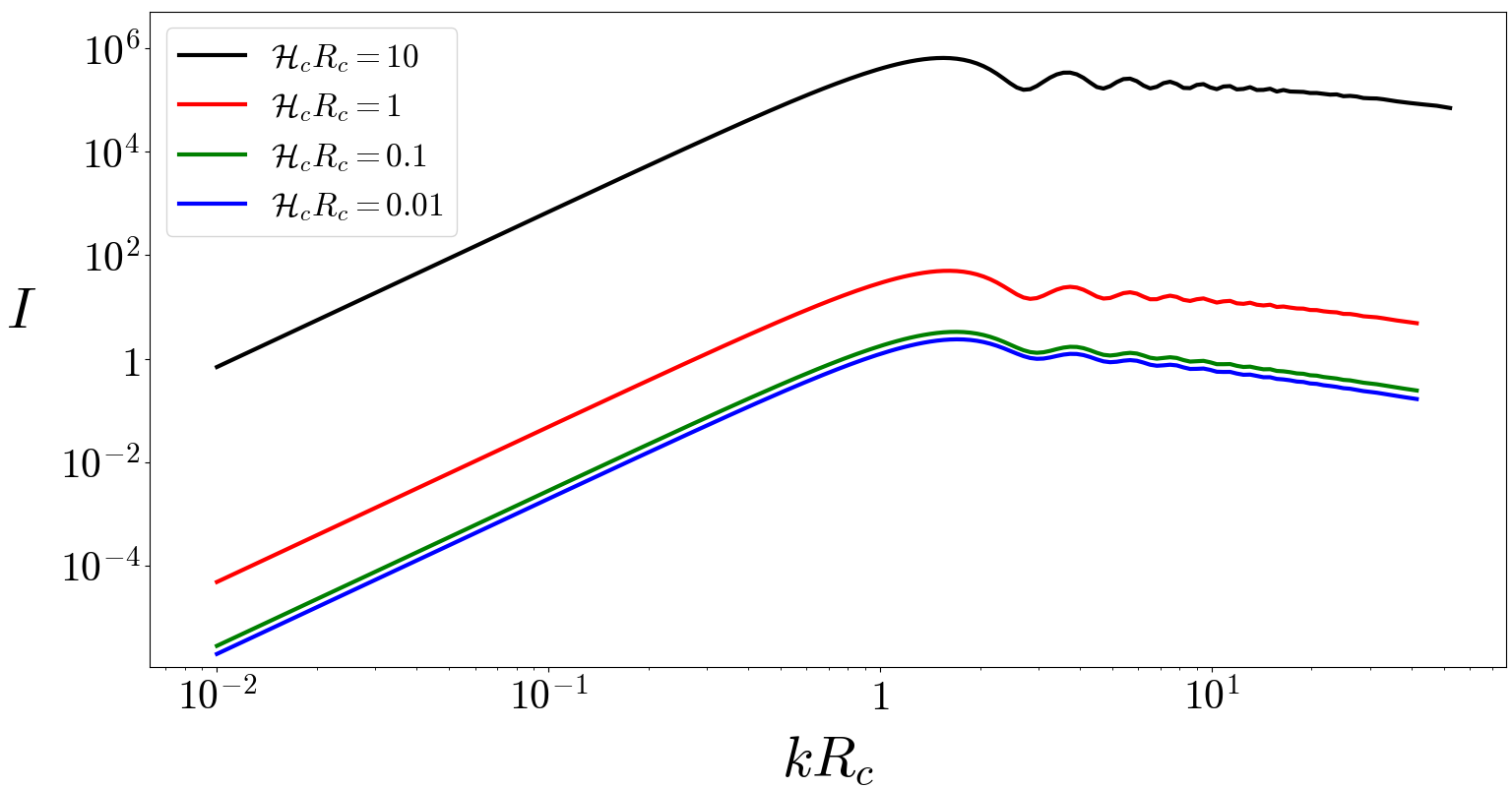}
    \caption{The function $I(kR_c)$~\eqref{eqn:function I} for various values of $\mathcal{H}_cR_c$. It is proportional to the GW spectrum today.}
    \label{fig:Ivuc}
\end{figure}

The plot of $I$ as a function of $kR_c$ for some chosen fixed values of the parameter $\mathcal{H}_cR_c$
is shown in Figure~\ref{fig:Ivuc}. 
For a generic FLRW spacetime, when $k < \mathcal{H}$ the different parts of the source are uncorrelated as they are separated by more than a Hubble length. This results in a scale invariant power spectrum for the source $\mathcal{P}_{\Pi}$ that in turn leads to a $k^3$ behaviour in $h^2 \Omega_\text{GW}$~\cite{Caprini:2018mtu}. However, as we remarked earlier, in the case of radiation background, the expansion term $d^2a/d\eta^2$ drops out of the equation of motion for the metric perturbations~\eqref{equ:htildeij}. Therefore the only scale arises from the radius of bubbles in the source term. Interestingly, in the case of super-Hubble bubbles, this implies that the $k^3$ behaviour only applies to scales longer than the bubble size, and does not extend all the way down to the Hubble length i.e. $k < R^{-1}_c < \mathcal{H}_c$. 

On the other hand, for $k \gtrsim R^{-1}_c$, the spectrum should possess a negative slope in order to preserve the area under the curve for the total gravitational radiation to remain finite~\cite{Caprini:2018mtu}. Combining the small and large $k$ behaviour, we expect the GW spectrum to possess a peak, as is evident from Figure~\ref{fig:Ivuc}. We estimate the peak frequency $k^\text{peak}$ by finding the first maxima of the function $I(kR_c)$ for a fixed $\mathcal{H}_cR_c$.

\begin{figure}
    \centering
    \includegraphics[scale=0.3]{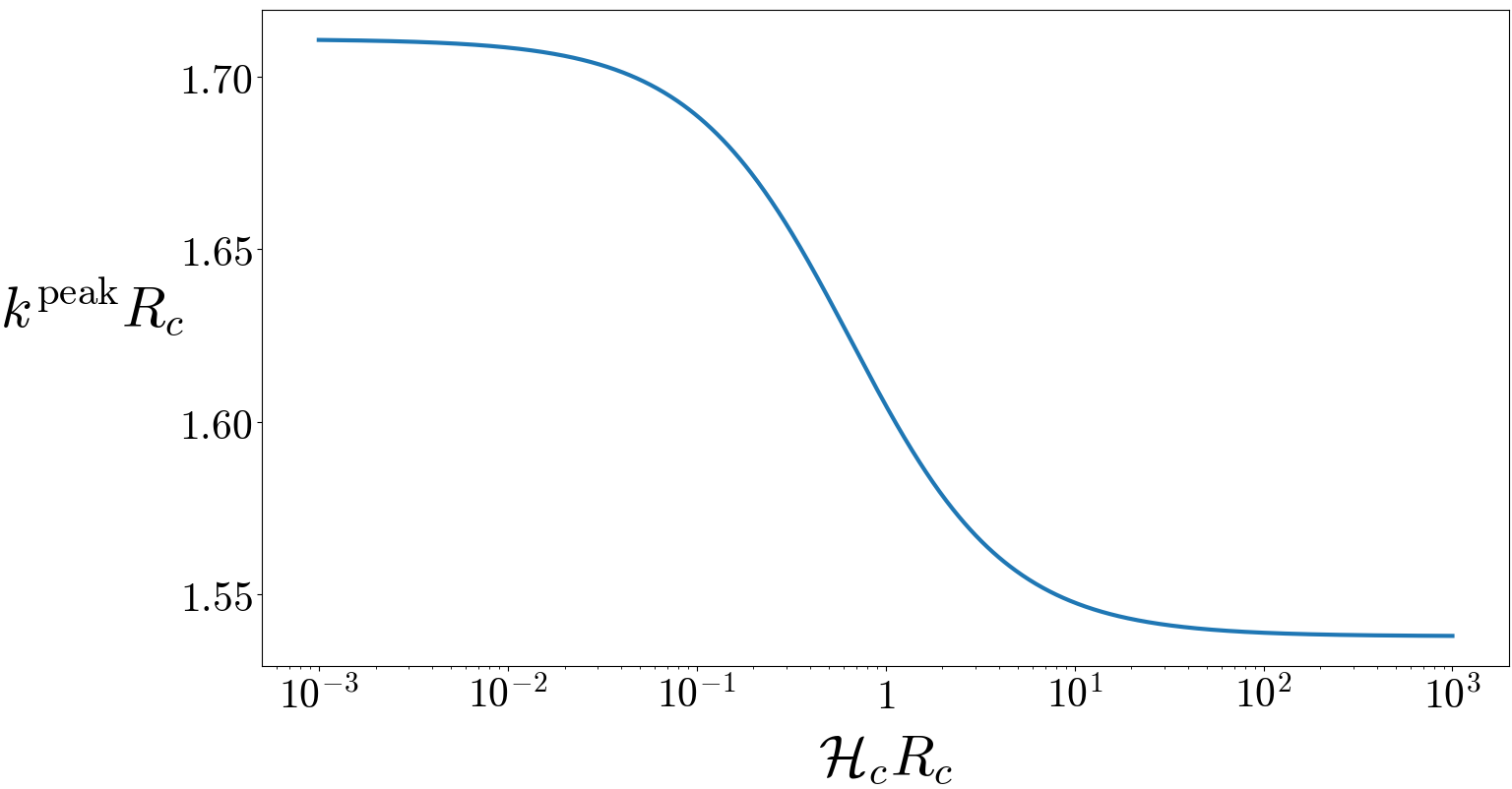}
    \caption{Plot of peak comoving frequency $k^\text{peak}R_c$ as a function of physical bubble size at collision $\mathcal{H}_cR_c$. We can separate the plot into two parameter regions, $\mathcal{H}_cR_c \ll 1$ and $\mathcal{H}_cR_c \gg 1$. The maximum of $I(kR_c)$~\eqref{eqn:function I} lies at $k^\text{peak}R_c \sim 1$ for both regions as expected (see text). However because of an additional cubic term in $u=R/R_c$ that becomes significant when $\mathcal{H}_cR_c \gg 1$, the peak $k^\text{peak}R_c$ shifts like a step function around $\mathcal{H}_cR_c \sim 1$.}
    \label{fig:kpeakRcvHcrc}
\end{figure}

The plot of peak frequency as a function of $\mathcal{H}_cR_c$ is shown in Figure~\ref{fig:kpeakRcvHcrc}. 
Since $I$ is a function of a single variable $kR_c$ for fixed $\mathcal{H}_cR_c$, one expects the peak amplitude to occur at $k^\text{peak}R_c \sim 1$. This is evident from Figure~\ref{fig:kpeakRcvHcrc}. There is a small step-function like shift in $k^\text{peak}R_c$ because the $u=R/R_c$-dependence of integral in $I(kR_c)$~\eqref{eqn:function I} changes between the parameter regions where $\mathcal{H}_cR_c \ll 1$ and $\mathcal{H}_cR_c \gg 1$. However, the conclusion remains that $k^\text{peak}R_c \approx 1.6$ for all values of $\mathcal{H}_cR_c$.

This gives the peak physical gravitational wave frequency today as~\eqref{eqn:physical and comoving frequency}
\begin{align}
\label{equ:fpeak}
    f^\text{peak} = \frac{1}{2\pi} \frac{k^\text{peak}}{a_0} \approx  \frac{0.25}{a_0R_c}. 
\end{align}
This result is the same as the one expected for conventional sub-Hubble bubble collisions~\cite{Caprini:2018mtu}. However, since more time elapses between nucleation and collision for super-Hubble bubbles, the comoving radius $R_c$ is able to grow significantly. Hence we are led to the generic conclusion that gravitational wave signal from super-Hubble bubbles peaks at a lower frequency than their conventional sub-Hubble counterpart when comparing the same start time at bubble nucleation.

It is also instructive to express the peak frequency in terms of the parameter $\mathcal{H}_cR_c$ as
\begin{align}\label{eqn:result for peak frequency}
    f^\text{peak} \approx 0.25 H_0 \frac{a_cH_c}{a_0H_0} \frac{1}{\mathcal{H}_cR_c}
    \approx 4\times 10^{-8}\,\text{Hz}\,\frac{T_c}{\text{GeV}}\frac{1}{\mathcal{H}_cR_c},
\end{align}
Comparison with eq.~(\ref{equ:fpeakthumb}) shows that for collisions happening at the same time $\eta_c$, 
super-Hubble bubbles lead to lower peak frequency than sub-Hubble bubbles due to the presence of the $(\mathcal{H}_cR_c)^{-1}$ factor in the above expression. 

Considering specifically a scenario with nucleation taking place $N_n$ e-folds before the end of inflation and using eq.~\eqref{eq:ellndef}, the peak frequency in eq.~\eqref{eqn:result for peak frequency} becomes
\begin{align}
    f^\text{peak} \approx 0.50 H_0 \frac{e^{N_*-N_n}}{\ell_n},
\end{align}
where we have also used the relation
\begin{align}
    \frac{a_\text{end}H_\text{end}}{a_0H_0} = e^{N_*},
\end{align}
with $N_*$ being the number of efolds before the end of inflation when the largest currently observable scale left the horizon. In principle, $N_*$ is dependent on the full cosmological evolution from inflation until the present. However, assuming instantaneous reheating followed by standard $\Lambda$CDM evolution, it can typically be approximated in terms of the energy scale of inflation $V^{1/4}_\text{inf}$ as~\cite{Markkanen:2018pdo}
\begin{align}\label{eqn:Nstar}
    N_* \approx 60 + \ln\left(\frac{V^{1/4}_\text{inf}}{10^{16} \text{GeV}}\right).
\end{align}
Thus we obtain the peak frequency for GW arising from bubbles that nucleate during inflation but collide during radiation to be 
\begin{align}\label{eqn:peak GW frequency from inflationary bubbles}
     f^\text{peak} \approx  (10^8 \text{Hz})\left(\frac{V^{1/4}_\text{inf}}{10^{16} \text{GeV}}\right)\frac{e^{-N_n}}{\ell_n}. 
\end{align}
In the conventional sub-Hubble collision scenario $(\mathcal{H}_cR_c \ll 1)$, the last factor $(e^{N_n} \ell_n)^{-1}$ in eq.~(\ref{eqn:peak GW frequency from inflationary bubbles}) would not be present, and therefore inflation-scale transitions would typically produce gravitational waves with MHz frequencies. In contrast, this factor results in lowering the frequency both in the supercooled $(\mathcal{H}_cR_c \approx 1)$ as well as the super-Hubble bubble collision $(\mathcal{H}_c R_c \gg 1)$ scenarios. In the supercooled case, because $\ell_n \gg 1$, the frequency is lowered even if $N_n$ lies very close to the end of inflation. On the other hand, for super-Hubble bubbles, we have $\ell_n \sim 2$, which implies that $N_n$ has to be relatively large to give rise to a low frequency signal. For example with $V^{1/4}_\text{inf} \approx 10^{15}$ GeV, typically $N_n \sim 30$ gives rise to nHz frequencies, $N_n \sim 20$ gives rise to mHz frequencies while $N_n \sim 10$ gives rise to frequencies in the hundreds of Hertz.   

\begin{figure}
    \centering
    \includegraphics[scale=0.3]{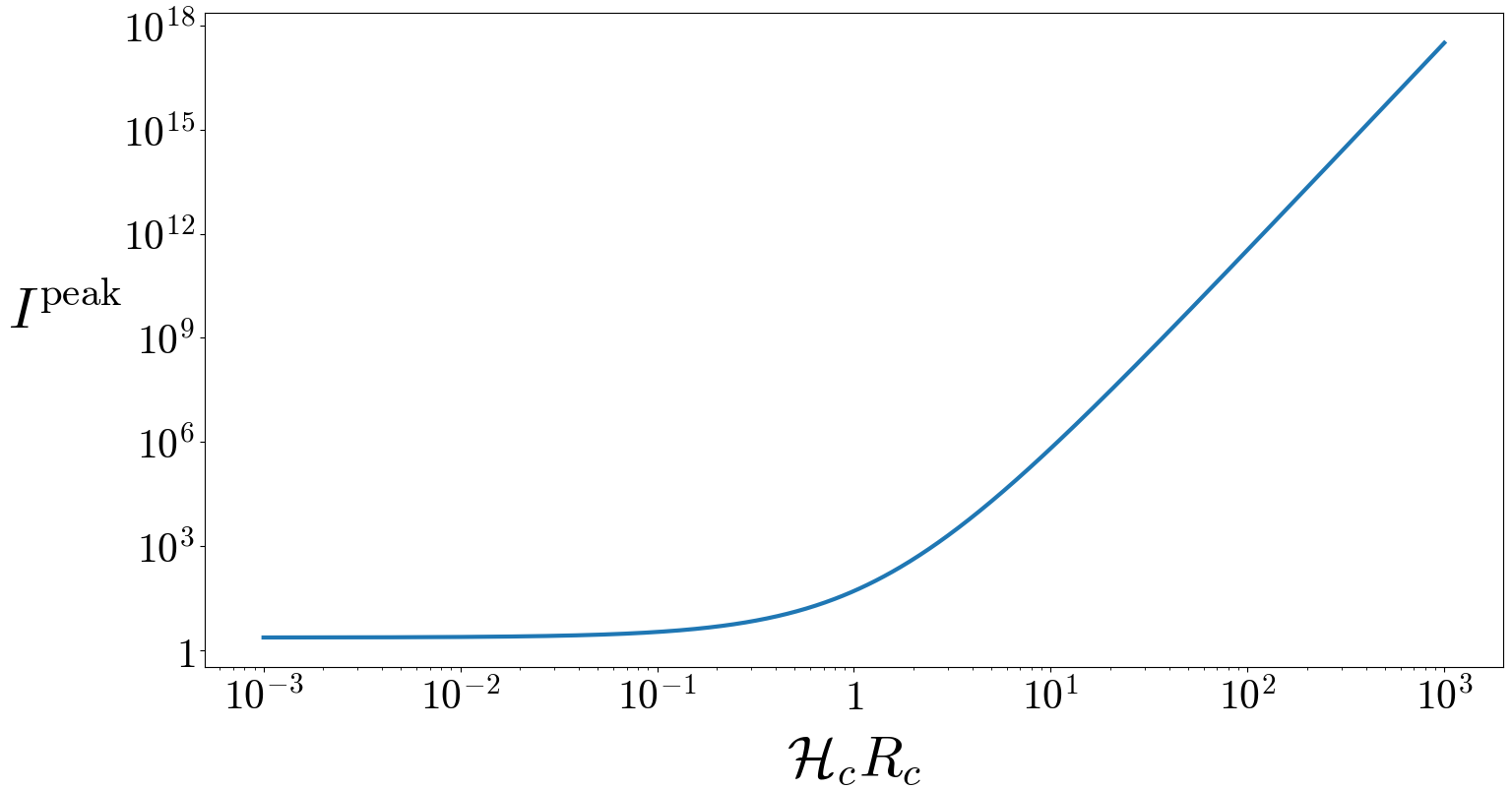}
    \caption{The peak value of the function $I$ defined in~\eqref{eqn:function I}, as a function of the physical bubble size at collision $\mathcal{H}_cR_c$.}
    \label{fig:IpeakvrH}
\end{figure}

From the peak frequency $k^\text{peak}$, it is straightforward to obtain the peak amplitude 
\begin{align}\label{eqn:peak amplitude}
  h^2\Omega_\text{GW}(k^\text{peak})  \sim 10^{-7} \frac{\Delta\rho^2}{\rho^2_c} (\mathcal{H}_cR_c)^2~I^\text{peak}(\mathcal{H}_c R_c)
\end{align}
with $I^\text{peak}(\mathcal{H}_cR_c) \equiv I(k^\text{peak}R_c,\mathcal{H}_cR_c)$. As a function of the parameter $\mathcal{H}_cR_c$, it is shown in Figure~\ref{fig:IpeakvrH}. We find that $I^\text{peak}(\mathcal{H}_cR_c \ll 1) \approx 2.3$ for sub-Hubble bubbles, $I^\text{peak}(\mathcal{H}_cR_c = 1) \approx 50$ for Hubble-sized bubbles and $I^\text{peak}(\mathcal{H}_cR_c  \gg 1) \approx 0.32 (\mathcal{H}_cR_c)^6$ for super-Hubble bubbles.  
Therefore, the peak amplitude for the different bubble sizes is
\begin{numcases}{h^2\Omega_\text{GW}(k^\text{peak}) \sim}
10^{-7} \frac{\Delta\rho^2}{\rho^2_c} (\mathcal{H}_cR_c)^2 
& for~$\mathcal{H}_cR_c \ll 1$, \label{eqn:Small bubble peak amplitude}\\
10^{-6} \frac{\Delta\rho^2}{\rho^2_c} & for~$\mathcal{H}_cR_c \sim 1$, \label{eqn:peakOmegaGWforHcrc1} \\
10^{-8} \frac{\Delta\rho^2}{\rho^2_c} (\mathcal{H}_cR_c)^8   & for~$\mathcal{H}_cR_c \gg 1$.\label{eqn:Big bubble peak amplitude}
\end{numcases}

Eq.~\eqref{eqn:Small bubble peak amplitude} underestimates the conventional expression for sub-Hubble bubbles prevalent in the literature~\cite{Kamionkowski:1993fg} by an order of magnitude. It is to be expected~\cite{Kosowsky:1992vn} since we are neglecting the collisions between three or more bubbles.

Eq.~\eqref{eqn:Big bubble peak amplitude} is the peak amplitude for super-Hubble bubbles. When comparing collisions which start at the same time $\eta_c$, it shows that the peak amplitude from super-Hubble bubbles is greatly enhanced as compared to Hubble-sized bubbles~\eqref{eqn:peakOmegaGWforHcrc1}.
However, the larger the bubbles, the longer their collision takes, and the more the Universe expands during it. Therefore, it is instructive to also compare collisions that complete at the same time $\eta_f$.
The amount of expansion during the collision is given by
\begin{align}
    \frac{a_f}{a_c} = 1 + \left(\frac{R_f}{R_c} -1  \right)   \mathcal{H}_cR_c
    = 1 + \left(u_f -1  \right)   \mathcal{H}_cR_c,
\end{align}
where $u_f$ is the cutoff parameter defined in eq.~(\ref{eqn:function I}).
Thus, written in terms of the energy density at the final time $\rho_f$, the peak amplitude~\eqref{eqn:resultOmegaGW} becomes
\begin{align}
\label{equ:GWrhof}
    h^2\Omega^\text{peak}_\text{GW} \sim 10^{-7} \frac{\Delta \rho^2}{\rho^2_f} ~\Xi(k^\text{peak}R_c, \mathcal{H}_cR_c,u_f) ,
\end{align}
with
\begin{align}\label{eqn:Xidefn}
    \Xi \equiv \frac{(\mathcal{H}_cR_c)^2}{\left(1 + \left(u_f -1  \right)   \mathcal{H}_cR_c\right)^8}~I(k^\text{peak}R_c,\mathcal{H}_cR_c).
\end{align}

\begin{figure}
    \centering
    \includegraphics[scale=0.3]{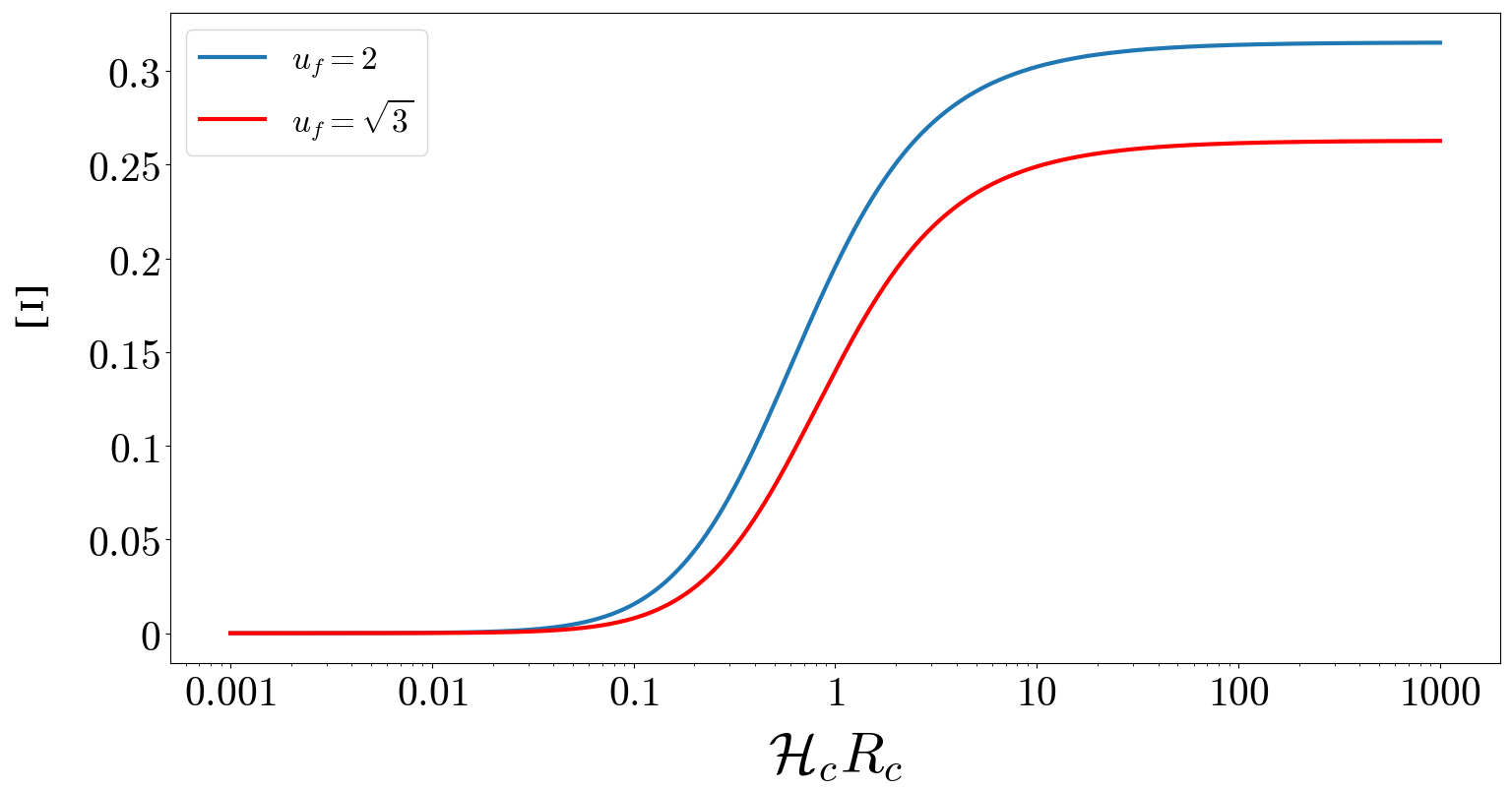}
    \caption{The factor $\Xi$~\eqref{eqn:Xidefn} as a function of $\mathcal{H}_cR_c$ for two different values of the cutoff parameter $u_f$.}
    \label{fig:XivHcrc}
\end{figure}

The plot of $\Xi$ as a function of $\mathcal{H}_cR_c$ is shown in Figure~\ref{fig:XivHcrc}. The function $I(k^\text{peak}R_c,\mathcal{H}_cR_c)$ depends on the choice of $u_f$, and as the plot shows, this largely cancels against the $u_f$-dependence of the denominator. Overall, the relative difference between the two choices shown is less than $20\%$.

Furthermore, it can be seen from Figure~\ref{fig:XivHcrc} that for large $\mathcal{H}_cR_c$, $\Xi$ approaches a constant $\approx 0.3$. 
This means that when comparing collisions completing at the same time, the amplitude of gravitational waves from super-Hubble bubbles with $\mathcal{H}_cR_c\gg 1$ is enhanced by only a factor of $\approx 2$ relative to Hubble-sized bubbles with $\mathcal{H}_cR_c\approx 1$.
Hence the peak amplitude from Hubble-sized as well as super-Hubble bubbles can be expressed as 
\begin{align}\label{eqn:peak amplitude at final time}
    h^2\Omega^\text{peak}_\text{GW} \sim 10^{-8} \frac{\Delta\rho^2}{\rho^2_f}~~~\text{for}~\mathcal{H}_cR_c \gtrsim 1.  
\end{align}
From the assumption~(\ref{eqn:kappa defn}), we conclude that we can only reliably describe the collision with our approximations if $h^2 \Omega_\text{GW}(k^\text{peak}) \lesssim 10^{-8}$.

Eqs.~(\ref{eqn:result for peak frequency}) and (\ref{eqn:Big bubble peak amplitude}) are our main result. They can be related to each other with the help of the radiation identities  $a^2_cH_c = a^2_\text{end} H_\text{end}$ and $a^4_c\rho_c = a^4_\text{end}\rho_\text{end}$ as
\begin{align}\label{eqn:relation between peak amplitude and frequency}
   h^2\Omega_\text{GW}(k^\text{peak})   \sim  10^{-13} \frac{\Delta\rho^2}{\rho^2_\text{end}}\left(\frac{a_\text{end}}{a_0}\right)^8 \left(\frac{H_\text{end}}{f^\text{peak}}\right)^8. 
\end{align}
Thus we find that there is a power law dependence between the peak amplitude and peak frequency for super-Hubble bubbles. Note that apart from the term $\Delta \rho$ in the above expression which depends on the transition, all other factors are fixed by the inflationary evolution.

In the next section, we illustrate with the help of a specific model, how super-Hubble bubble collisions can lead to a detectable signal at current and planned gravitational wave experiments.

\section{Illustrative model}
\label{sec:Illustrative model}
Consider a theory with two scalar fields, the inflaton $\varphi$ and a transition field $\sigma$. The total potential $W(\varphi,\sigma)$ is composed of the inflationary part $V(\varphi)$, transition part $U(\sigma)$ and a coupling between them that is quadratic in $\sigma$, 
\begin{align}\label{eqn:inflation totPot}
    W(\varphi,\sigma) = V(\varphi) + U(\sigma) - g(\varphi)\frac{\sigma^2}{2},
\end{align}
where $g(\varphi)$ is a positive function of the inflaton field, $g(\varphi) > 0$.

We assume that the transition sector including the coupling term remains subdominant to the inflationary sector i.e. 
\begin{align}
    V(\varphi) \gg U(\sigma) - g(\varphi) \frac{\sigma^2}{2}.
\end{align}
Therefore we can study the background evolution purely based on the inflationary potential $V(\varphi)$. As an inflationary model that lies well-within current observational bounds, we choose the inflaton part to be of the alpha-attractor T-model form~\cite{Kallosh:2013hoa,Kallosh:2013yoa}
\begin{align}\label{eqn:inf pot}
    V(\varphi) = \lambda M^4_P \tanh^4\left(\frac{\varphi}{\sqrt{6} M_P}\right).
\end{align}
Taking inflation to end when the slow-roll parameter 
\begin{align}
    \epsilon \equiv \frac{M^2_P}{2} \left(\frac{V_{,\varphi}}{V}\right)^2 = 1 \implies \varphi_\text{end} = 1.93M_P.
\end{align}
We calculate the number of efolds when the largest currently observed scale left the horizon $N_*$ using eq.~\eqref{eqn:Nstar} by taking $V_\text{inf} = V(\varphi_\text{end})$. Then using
\begin{align}
    N(\varphi) = \frac{1}{M^2_P} \int_{\varphi_\text{end}}^{\varphi} \frac{V}{V_{,\varphi}} d\varphi
\end{align}
in slow-roll, we obtain the inflaton value $\varphi_*$ corresponding to $N_*$ efolds. Using this, the curvature power spectrum in slow--roll can be obtained as
\begin{align}
    P_\zeta = \frac{1}{24 \pi^2 M^4_P} \frac{V}{\epsilon}\Bigg|_{\varphi=\varphi_*}.
\end{align}
Thus to generate the correct observed curvature power spectrum of $P_\zeta = 2.1 \times 10^{-9}$~\cite{Tristram:2023haj} we find that $N_* = 59.33$ at $\varphi_* = 7.07M_P$ and obtain $\lambda = 1.05 \times 10^{-10}$. We can also verify that the tensor-to-scalar ratio for this model is 
\begin{align}
    r = 16 \epsilon \Big|_{\varphi=\varphi_*} = 0.0033,
\end{align}
which lies well-within the observable bound of $r < 0.032$~\cite{Tristram:2021tvh}. Thus this model is observationally viable. Assuming that all the energy density at end of inflation is transferred to radiation via instant reheating, the reheat temperature in this model is 
\begin{equation}
    \label{equ:Treheat}
T_\text{reheat} \sim 10^{15}~\text{GeV}.
\end{equation}

The transition sector is taken to be of a simple quartic form,
\begin{align}
    U(\sigma) = \frac{\lambda'}{4!} \sigma^4 + \frac{g'}{3!} \sigma^3 - \frac{\mu^2}{2!} \sigma^2,
\end{align}
where the parameters $\lambda'$, $g'$ and $\mu$ are all assumed positive. For a given fixed value of $\varphi$, the potential has a barrier at $\sigma_\text{bar}=0$ separating two minima. We denote the positive value by $\sigma_\text{FV}$ since it has a higher potential and thus corresponds to a false vacuum state; the negative one is the true vacuum, denoted by $\sigma_\text{TV}$. Without the cubic term, there would be a $\sigma\rightarrow -\sigma$ symmetry. Therefore small values of the cubic coupling $g'$ are technically natural~\cite{tHooft:1979rat}. Hence we generally assume
\begin{align}\label{eqn:gprimevalue}
    g'^2\ll \lambda' \mu^2_\text{eff}~.
\end{align}
Combined with the inflaton coupling term, we have an effective mass for the transition field $\sigma$,
\begin{align}\label{eqn:mu effective}
    \mu^2_\text{eff}(\varphi) \equiv -\left.\frac{\partial^2 W}{\partial\sigma^2}\right|_{\sigma=\sigma_\text{bar}} =  \mu^2 + g(\varphi).
\end{align}
Owing to eq.~\eqref{eqn:gprimevalue}, the false and true vacua are nearly degenerate and given by $\sigma_{\text{FV}/\text{TV}}\approx \pm \sqrt{6\mu_\text{eff}^2/\lambda'}$.

The bubble nucleation rate as a function of the inflaton field is given by~\cite{Hawking:1981fz}
\begin{align}
    \label{equ:HMrate}
    \Gamma(\varphi) \sim H(\varphi)^4 e^{-B_\text{HM}(\varphi)}
\end{align}
where
the Hawking-Moss instanton action is
\begin{align}\label{eqn:HM action}
    B_\text{HM}(\varphi) = \frac{8 \pi^2}{3H(\varphi)^4} \bigl(W(\varphi,\sigma_\text{bar}) - W(\varphi,\sigma_{\rm{FV}})\bigr)
   \approx \frac{4\pi^2}{\lambda'} \frac{\mu_\text{eff}(\varphi)^4}{ H(\varphi)^4}. 
\end{align}
We assume that $\sigma$ is initially in its false vacuum state. As the background $\varphi$ evolves to lower values towards the end of inflation, bubble nucleation starts to become more favourable. With the alpha-attractor background~\eqref{eqn:inf pot} that remains almost flat throughout slow-roll inflation, it is straightforward to show that for the number of bubbles to peak~\eqref{eqn:gamma definition} at $\varphi_n$ requires $g_{,\varphi_n} < 0$. We choose 
\begin{align}
    g(\varphi) = g^2_I M^2_P\tanh^2\left(\frac{\varphi}{M_P} - c\right).
\end{align}
For this choice of the coupling function, $g_{,\varphi_n} < 0$ when $\varphi_n < c M_P$. Thus an appropriate choice of the parameter $c$ ensures that maximum number of bubbles are nucleated towards the end of inflation, within the observable number of efolds.

To estimate the peak amplitude of gravitational waves produced, we shall assume that the regions inside and outside the bubble continue to evolve after the end of inflation and hence the value of the inflaton field has settled down at the bottom of its potential to $\varphi=0$ by the time bubbles collide in the radiation era. Therefore,
\begin{align}
    \Delta\rho = W(0,\sigma_{\rm{FV}}) - W(0,\sigma_{\rm{TV}}) \approx 2\sqrt{6} g' \left(\frac{\mu^2 + g^2_I M^2_P \tanh^2c}{\lambda'} \right)^{3/2}.
\end{align} 
The peak amplitude of gravitational waves~\eqref{eqn:Big bubble peak amplitude} is thus proportional to $g'^2$.

The assumption $\Delta \rho/\rho < 1$~\eqref{eqn:kappa defn} ensures that the false vacuum energy density does not dominate the Universe at any point between bubble nucleation and end of collision~\cite{Ellis:2018mja}. This is satisfied in our case due to eq.~\eqref{eqn:gprimevalue}.

We consider two sets of parameters A and B as specified in Table~\ref{tab:model parameters}. For both sets of parameters, we have checked the conditions necessary for bubble nucleation to proceed through Hawking-Moss instantons~\cite{Balek:2003uu}, namely that $\mu_\text{eff} \lesssim 2H$ and $B_\text{HM}> 1$ at the nucleation time $N_n$. This ensures that the Hawking-Moss nucleation rate~(\ref{equ:HMrate}) is valid. 

\begin{table}[t]
    \centering
    \def\arraystretch{1.2}
    \begin{tabular}{|c|c|c|c|c|c|c|}
    \hline
        Set & $g'$ & $c$  &  $g^\text{max}_I$  & $(\mathcal{H}_cR_c)^\text{max}$ \\ \hline

        A & $20$ MeV & $5.45$ & $1.45 \times 10^{-5}$ &  $5.96 \times 10^5$\\

        B & $8 \times 10^{-10}$ eV & $6$ & $2.38 \times 10^{-5}$ & $5.32 \times 10^{9}$ \\ 
        
    \hline
     \end{tabular}
    \caption{Parameter sets A and B for the illustrative model~\eqref{eqn:inflation totPot}. $\lambda'=1$ and $\mu = 4 \times 10^{12}~\text{GeV}$ for both sets. The maximum value of the bubble radius at collision $(\mathcal{H}_cR_c)^\text{max}$ occurs at $g^\text{max}_I$.}
    \label{tab:model parameters}
\end{table}

Once all other parameters are fixed, the maximum value of the parameter $g_I$ is restricted by the requirement that bubble collisions occur after inflation ends~\eqref{eqn:ellnbound}. We denote this value by $g^\text{max}_I$. It is also at this value that the bubble radius at collision $\mathcal{H}_cR_c$~\eqref{eqn:infHcrc} is the largest. This $(\mathcal{H}_cR_c)^\text{max}$ is specified for each of the parameter sets in Table~\ref{tab:model parameters}. 

\begin{figure}
    \centering
    \includegraphics[scale=0.3]{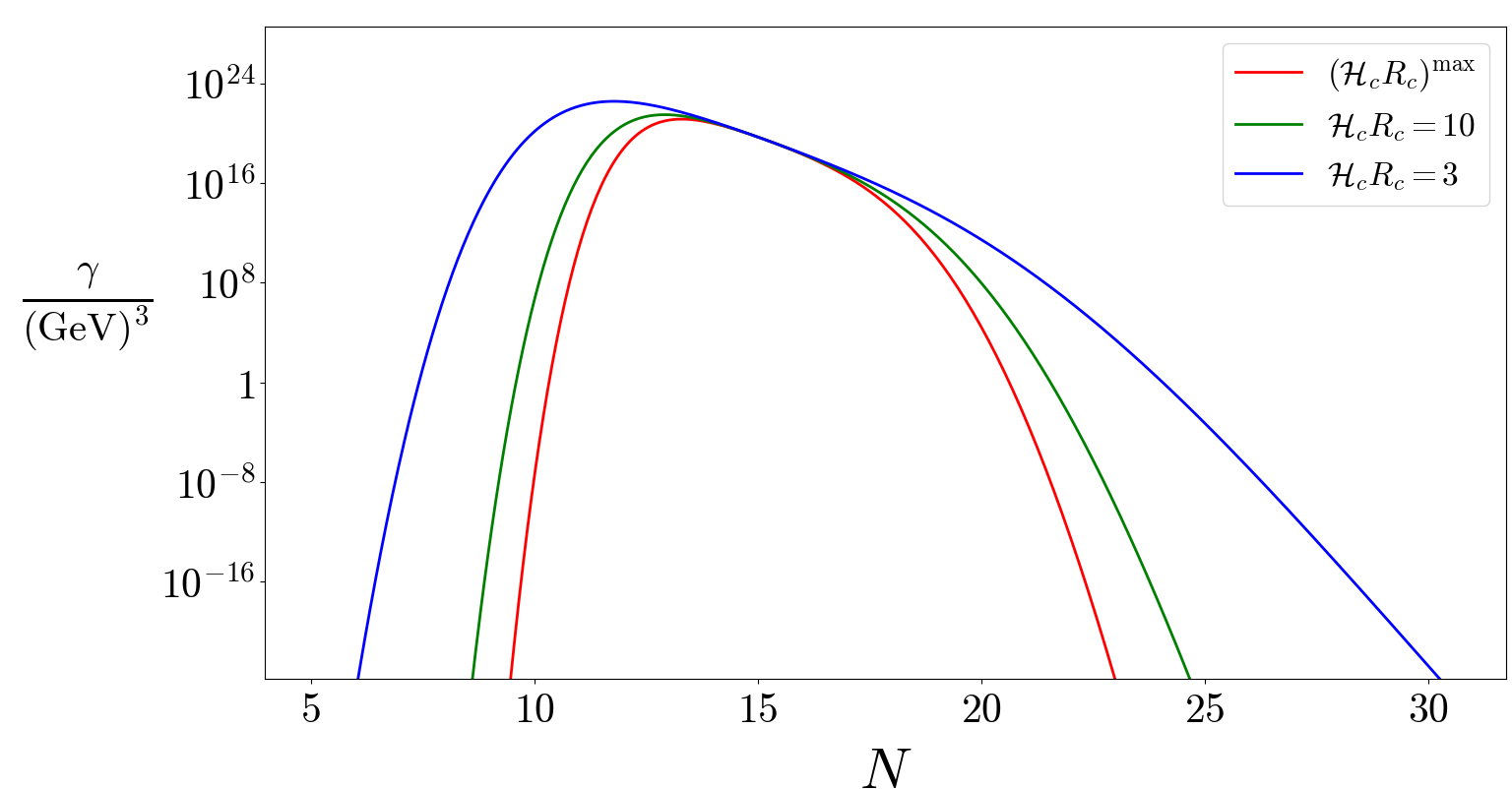}
    \caption{The function $\gamma(N)$ in eq.~\eqref{eqn:gamma definition} for different values of $g_I$ corresponding to the bubble radii $\mathcal{H}_cR_c = 3,~10,~ (\mathcal{H}_cR_c)^\text{max}$ for parameter set A in Table~\ref{tab:model parameters}. Its peak determines the number of efolds before the end of inflation $N_n$ when the maximum number of bubbles are nucleated.}
    \label{fig:nbubvN}
\end{figure}

As $g_I$ decreases from its maximum value, the maximum number of bubbles are nucleated closer and closer to the end of inflation.  This can be seen from Figure~\ref{fig:nbubvN}. As $N_n$ decreases, $\mu_\text{eff}$ starts to increase because of its $\tanh^2$ dependence. This results in the separation between nucleation sites $\ell_n$~\eqref{eqn:ell_n in terms of Gamma} to grow. The combination of $N_n$ decreasing and $\ell_n$ increasing in turn causes the bubble radius at collision~\eqref{eqn:infHcrc} to decrease from its maximum value.  

Significantly super-Hubble bubble sizes at collision arise only for a narrow range of $g_I$ values. Essentially because the collisions should take place not long after the end of inflation for bubbles to remain super-Hubble~\eqref{eqn:bubble radius evolution}. For example, the bubble radius is more than three times the Hubble length, $\mathcal{H}_cR_c \geq 3$ within the range $g_I \in [9.01 \times 10^{-6},~1.45\times 10^{-5}]$ for parameter set A and $g_I \in [1.52 \times 10^{-5},~2.38  \times 10^{-5}]$ for parameter set B. 

\begin{figure}[t]
    \centering
    \includegraphics[width = \textwidth]{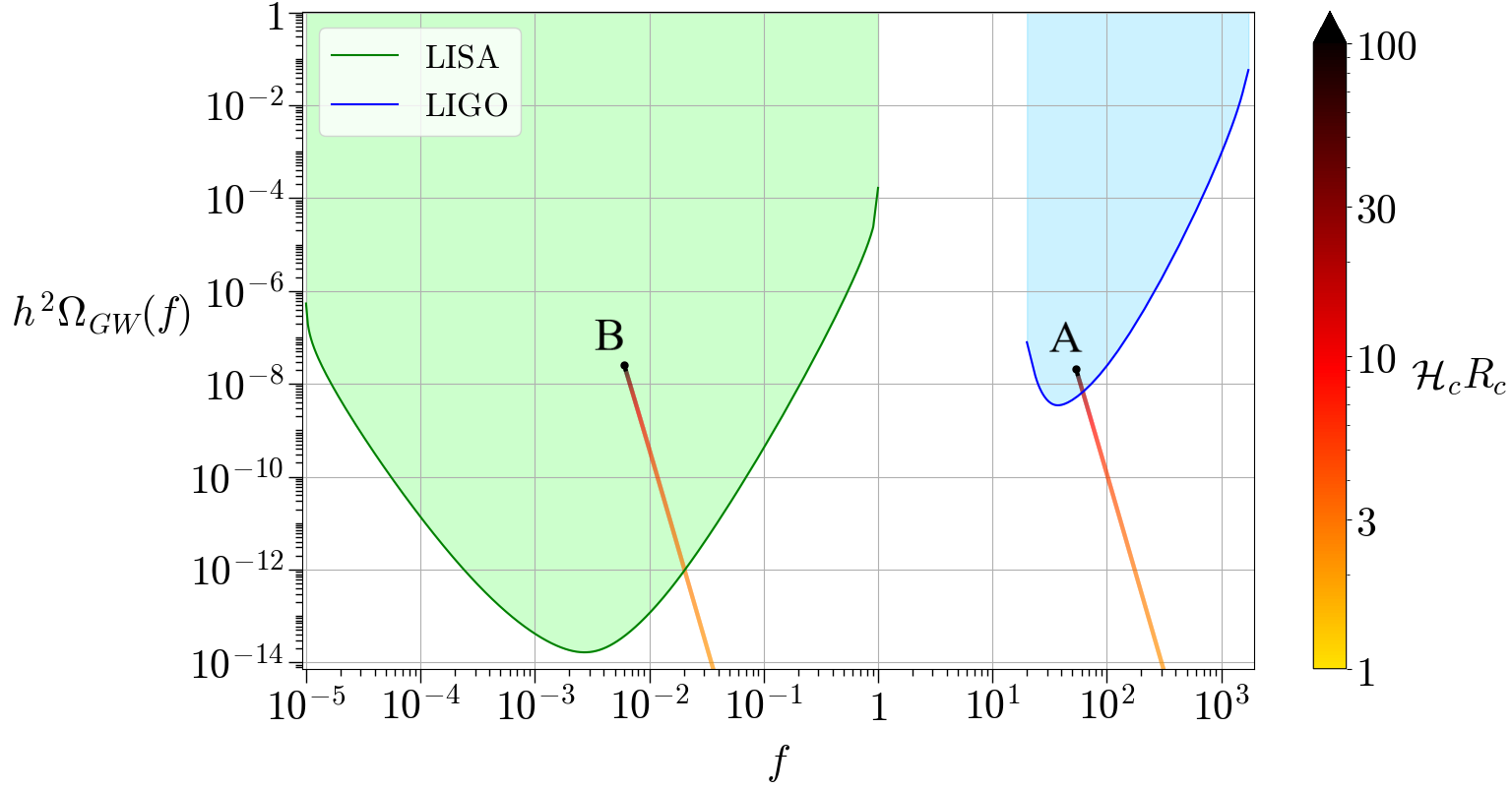}
    \caption{The peak GW frequency and amplitude corresponding to the parameter sets A and B in Table~\ref{tab:model parameters} and where they lie relative to the sensitivity windows
    of LISA~\cite{Schmitz:2020syl} and LIGO~\cite{KAGRA:2021kbb} experiments as the coupling $g_I$ is varied. The maximum values at $g_I = g^\text{max}_I$ are indicated with black dots. The different shades indicate the values of bubble radius at collision $\mathcal{H}_cR_c$.\\
    The plot for LISA is the Power-law Integrated (PI) sensitivity curve for first-order phase transitions~\cite{Schmitz:2020syl} available at \href{https://zenodo.org/records/3689582}{https://zenodo.org/records/3689582}. The LIGO plot is the Power-law Integrated (PI) sensitivity curve at 1 sigma, combining O1-O2-O3 runs of advanced LIGO and advanced Virgo. It is taken from the data accompanying ref.~\cite{KAGRA:2021kbb} publicly available at \href{https://dcc.ligo.org/G2001287/public}{https://dcc.ligo.org/G2001287/public}.}
\label{fig:GWsensitivities}
\end{figure}

Figure~\ref{fig:GWsensitivities} shows the peak GW frequency~\eqref{eqn:result for peak frequency} and amplitude~\eqref{eqn:Big bubble peak amplitude} as $g_I$ is decreased from its maximum value $g^\text{max}_I$ for the two different sets of parameters A and B respectively. The curves for both sets are straight lines on a log scale because of the power law dependence between the peak amplitude and frequency identified in eq.~\eqref{eqn:relation between peak amplitude and frequency}.

If we had not taken the curved spacetime effect due to super-Hubble bubbles into account i.e. if we ignored $I^\text{peak}$ in eq.~\eqref{eqn:peak amplitude} by setting it equal to unity, then the peak amplitude due to super-Hubble bubbles~\eqref{eqn:Big bubble peak amplitude} would be off by a factor of approximately $10^{36}$ and $10^{54}$ at the maximum bubble size $(\mathcal{H}_cR_c)^\text{max}$ of parameter sets A and B in Table~\ref{tab:model parameters} respectively. Crucially, by taking the effect into account, we find that the signal lies within the sensitivity windows of LIGO and LISA experiments.

Since LIGO has not observed any stochastic gravitational wave background, the region of parameter space of set A that lies within its
sensitivity range is likely disfavoured by existing data. However for set B, the parameter space that intersects with the forecasted LISA sensitivity range could still be detected in the near future.

These results show that super-Hubble bubble collisions taking place at very high energies can produce gravitational waves of essentially any frequency and amplitude depending on the model parameters. The parameter $g'$ was chosen to be small in order to satisfy the condition (\ref{eqn:kappa defn}), but even this is not a constraint imposed by physics but by the limitations of our approximations. Apart from this, the parameter values we used were generic, and therefore we expect the signal from super-Hubble bubble collisions in this model to cover a large range of the sensitivities of current and upcoming gravitational wave experiments. However, because the model is intended for illustration only, we do not carry out an analysis of its full parameter space.

\section{Discussion}
As the prospect of detecting gravitational waves from first-order phase transitions is becoming realistic in the near future, it becomes necessary to account for all possible mechanisms that yield a detectable signal. Through this article, we have shown how super-Hubble sized bubble collisions taking place after the end of inflation can be responsible for a signal within the sensitivity range of current and planned experiments.

By including curved spacetime effects, we estimated the stochastic gravitational wave spectrum~(\ref{eqn:resultOmegaGW}) from super-Hubble bubble collisions taking place during the radiation era and extracted its peak frequency~(\ref{eqn:result for peak frequency}) and amplitude~(\ref{eqn:peak amplitude}). Compared to Hubble-sized bubbles, the peak amplitude is significantly enhanced when comparing collisions beginning at the same time. This can be understood as a consequence of the expansion of the Universe during the collision. The peak frequency is determined by the comoving bubble radius, just like it is for smaller bubbles, and therefore it is lower for super-Hubble bubbles.

In order to carry out our calculation analytically i.e. without resort to numerical simulations, we have made some approximations
in addition to the standard ones~\cite{Kosowsky:1991ua} of linearised gravity and the envelope approximation.
The first approximation we make is only considering pairwise collisions between bubbles. This is able to provide a conservative underestimate of the total spectrum~\cite{Kosowsky:1992vn} which arises also from collisions between three and more bubbles. The second approximation 
is that the radii of bubbles at collision are equal. This is justified in the case where the decay rate is sharply-peaked at the nucleation threshold. The third approximation is that the different bubble pairs are uncorrelated, which allows us to average over their positions and orientations. An important open question is whether the enhancement~(\ref{eqn:Big bubble peak amplitude}) survives beyond these approximations.

Refs.~\cite{Zhong:2021hgo, Yamada:2025cfr} provide a more precise analytical treatment for the gravitational wave spectrum from bubble collisions in FLRW spacetimes by including multi-bubble collisions. But ultimately it is numerical simulations of the collision between multiple bubbles~\cite{Kosowsky:1992vn,Huber:2008hg,Child:2012qg,Cutting:2018tjt} that can provide the most accurate way to overcome the approximations stated above. This is a possible direction for future work. 

It would also be interesting to study how the super-Hubble nature of bubbles affects the gravitational waves generated by non-equilibrium processes that follow the collisions between them. Numerical simulations would be helpful in this direction as well.

We have focused on the peak frequency and amplitude of the GW signal as this information would be observed first at GW experiments, before the shape of the spectrum is measured. In this work, we have assumed instantaneous reheating to a radiation dominated era after inflation. For super-Hubble collisions taking place during the radiation era, we find that the characteristic shape of the GW spectrum would still be a broken power law just like for Hubble-sized or smaller bubbles. This is indicated in Figure~\ref{fig:Ivuc}. It is the case because the expansion scale $H^{-1}$ drops out of the metric equation of motion, leaving the radius of bubbles at collision as the only scale in the problem. 

It would be interesting to find out what the signal and its characteristic shape would look like for a different spacetime background where both the expansion scale $H^{-1}$ and radius of bubbles would play a role. For example, the period of reheating can be followed by or itself be an era of matter domination. Expanding the calculation for gravitational waves generated from super-Hubble bubble collisions to reheating or non-radiation backgrounds could be a direction for further work.

There are only a few ways of accessing inflationary physics via cosmological experiments. Of them, the tensor-to-scalar ratio and non-Gaussianity are proportional to the slow-roll parameters which are generically assumed to be small, making it difficult to distinguish between different inflationary models. This is where non-equilibrium phenomena such as preheating and phase transitions can help to enhance the observable signal. Through this article, we have shown that gravitational waves from super-Hubble bubble collisions just after the end of inflation are an additional way to access inflationary physics.

Not only is the GW signal from super-Hubble bubbles accessible at lower frequency experiments like LIGO, LISA and even PTA, but it will also be accessible at any planned high frequency experiments like the Einstein Telescope~\cite{ET:2025xjr}, essentially covering the full range of frequencies up to MHz. An additional advantage in terms of detection prospects for a GW signal from super-Hubble bubbles is that its amplitude is greatly enhanced as compared to other types of first-order phase transitions with smaller-sized bubble collisions that begin at the same time. 

In summary, we have demonstrated that the traditional notion of inflationary or GUT-scale phase transitions giving rise to high frequency gravitational waves does not hold true for super-Hubble bubbles. Instead, inflationary-scale first-order phase transitions with super-Hubble bubbles can generate LIGO/LISA-frequency gravitational waves with observable amplitudes.

\acknowledgments
We wish to thank Chiara Caprini and Jorinde van de Vis for helpful discussions.
PSG would like to thank Giorgio Mentasti for help with plotting GW sensitivity curves, the Physics Department at \'{E}cole Normale Sup\'{e}rieure (ENS) in Paris and acknowledge support from the French Embassy in the UK and the Government of Maharashtra, India. 
AR was supported by the UK Science and Technology Facilities Council (STFC) grant ST/X000575/1, a CERN Scientific Associate position, and a Visiting Fellowship at Wadham College, Oxford.

\bibliography{shortbib}
\bibliographystyle{JHEP}
\end{document}